\documentclass[preprint]{aastex}
\usepackage{emulateapj5}
\usepackage{apjfonts}
\usepackage{natbib}

\input psfig.tex

\def\aa{{A\&A}}

\def\aj{{AJ}}
\def\al{$\alpha$}
\def\bet{$\beta$}

\def\annrev{{ARA\&A}}
\def\apj{{ApJ}}
\def\apjs{{ApJS}}

\def\deg{$^{\circ}$}

\def\kms{km s$^{-1}$}
\def\lamb{$\lambda$}
\def\lax{{$\mathrel{\hbox{\rlap{\hbox{\lower4pt\hbox{$\sim$}}}\hbox{$<$}}}$}}
\def\gax{{$\mathrel{\hbox{\rlap{\hbox{\lower4pt\hbox{$\sim$}}}\hbox{$>$}}}$}}
\def\simlt{\lower.5ex\hbox{$\; \buildrel < \over \sim \;$}}
\def\simgt{\lower.5ex\hbox{$\; \buildrel > \over \sim \;$}}
\def\lum{ergs s$^{-1}$}
\def\mbh{{$M_{\rm BH}$}}

\def\mnras{{MNRAS}}

\def\percm2{cm$^{-2}$}
\def\peryr{yr$^{-1}$}

\def\solmass{$M_\odot$}

\def\hi{\ion{H}{1}}
\def\hii{\ion{H}{2}}
\def\oii{[\ion{O}{2}]}
\def\oiii{[\ion{O}{3}]}

\def\nii{[\ion{N}{2}]}

\def\sii{[\ion{S}{2}]}

\def\lbol{$L_{{\rm bol}}$}
\def\ledd{$L_{{\rm Edd}}$}

\def\sig{$\sigma_*$}

\def\msigma{$M_{\rm BH}-\sigma_*$}
\def\mhi{$M_{{\rm H~I}}$}

\def\vm{${\upsilon_m}$}

\def\vmsig{$\upsilon_m-\sigma_*$}
\def\mvm{$M_{\rm BH}-\upsilon_m$}

\slugcomment{To appear in {\it The Astrophysical Journal}.}
\shorttitle{\hi\ in ACTIVE GALAXIES}
\shortauthors{HO, DARLING, \& GREENE}

\begin{document}

\title{Properties of Active Galaxies Deduced from \hi\ Observations}

\author{Luis C. Ho\altaffilmark{1}, Jeremy Darling\altaffilmark{2}, and
Jenny E. Greene\altaffilmark{3,4}}

\altaffiltext{1}{The Observatories of the Carnegie Institution of Washington,
813 Santa Barbara St., Pasadena, CA 91101.}
\altaffiltext{2}{Center for Astrophysics and Space Astronomy, Department of
        Astrophysical and Planetary Sciences, University of Colorado, 389 UCB, 
        Boulder, CO 80309-0389.}
\altaffiltext{3}{Princeton Observatory, Peyton Hall, Princeton University, 
Princeton, NJ 08544-1001.}
\altaffiltext{4}{Hubble Fellow and Carnegie-Princeton Fellow.}

\begin{abstract}
We have completed a new survey for \hi\ emission for a large, well-defined 
sample of 154 nearby ($z$ \lax\ 0.1) galaxies with type~1 (broad-line) active 
galactic nuclei (AGNs).  We make use of the extensive database of \hi\ and 
optical parameters, presented in a companion paper, to perform a comprehensive 
appraisal of the cold gas content in active galaxies and to seek new 
strategies to investigate the global properties of the host galaxies and their 
relationship to their central black holes.  After excluding objects with 
kinematically anomalous line profiles, which occur with high frequency in the 
sample, we show that the black hole mass obeys a strong, roughly linear 
relation with the host galaxy's dynamical mass, calculated by combining the 
\hi\ line width and the optical size of the galaxy.  Black hole mass follows 
a looser, though still highly significant, correlation with the maximum 
rotation velocity of the galaxy, as expected from the known scaling between 
rotation velocity and central velocity dispersion.  Neither of these \hi-based 
correlations is as tight as the more familiar relations between black hole mass 
and bulge luminosity or velocity dispersion, but they offer the advantage of 
being insensitive to the glare of the nucleus and therefore are promising new 
tools for probing the host galaxies of both nearby and distant AGNs.  We 
present evidence for substantial ongoing black hole growth in the most actively 
accreting AGNs.  In these nearby systems, black hole growth appears to be 
delayed with respect to the assembly of the host galaxy but otherwise has 
left no detectable perturbation to its mass-to-light ratio, as judged from the 
Tully-Fisher relation, or its global gas content.  The host galaxies of type~1 
AGNs, including those luminous enough to qualify as quasars, are generally 
gas-rich systems, possessing a cold interstellar medium reservoir at least as 
abundant as that in inactive galaxies of the same morphological type.  This 
calls into question current implementations of AGN feedback in models of 
galaxy formation that predict strong cold gas depletion in unobscured AGNs.
\end{abstract}

\section{Introduction}

Dynamical studies of the central regions of nearby inactive galaxies have 
revealed that supermassive black holes (BHs; \mbh\ $\approx 10^6-10^9$ 
\solmass) are ubiquitous, and that their masses are strongly coupled to the 
host galaxy's bulge luminosity (Kormendy \& Richstone 1995; Magorrian et al. 
1998) and stellar velocity dispersion (\msigma\ relation: Gebhardt et al. 
2000a; Ferrarese \& Merritt 2000; Tremaine et al. 2002). The occupation 
fraction of central BHs in bulgeless or very late-type galaxies is poorly 
constrained, but at least some such systems have been found to contain BHs 
with masses as low as $\sim 10^5$ \solmass\ (Filippenko \& Ho 2003; Barth et 
al. 2004; Greene \& Ho 2004, 2007b, 2007c), which continue to obey the 
\msigma\ relation (Barth et al. 2005; Greene \& Ho 2006b).  These empirical 
correlations, which strongly suggest that BH growth is closely coupled to 
galaxy formation and evolution, have inspired considerable observational and 
theoretical attention in the last few years (see, e.g., reviews in Ho 2004a).  
BH growth, its observational manifestation as nuclear activity, and the 
consequences of feedback from active galactic nuclei (AGNs) are now widely 
viewed as unavoidable pieces of the overall puzzle of cosmological structural 
formation (e.g., Granato et al.  2004; Springel et al. 2005; Hopkins et al. 
2006).

The BH-bulge correlations beg several pressing, unanswered questions.  Do BHs
grow predominantly by radiatively efficient accretion  during the luminous (Yu
\& Tremaine 2002) or obscured (Fabian 1999) quasar phase, by low accretion
rates in moderately luminous AGNs at recent times (e.g., Cowie et al. 2003),
or by mergers (e.g., Yoo \& Miralda-Escud\'e 2004)?  Which came first, BH or
galaxy?  Must the growth of the BH and its host be finely synchronized so as
to preserve the small intrinsic scatter observed in the local BH-host scaling
relations?

These issues can be observationally tackled in two steps: first by extending
the BH-host scaling relations to {\it active}\ galaxies, wherein the BHs are
still growing, and second by probing the evolution of the scaling relations by
stepping back in redshift.  To this end, two ingredients are needed---BH
masses and host galaxy parameters.  Fortunately, \mbh\ can be estimated
readily in broad-line (type 1) AGNs, solely from basic spectroscopic data 
using the mass-luminosity-line width relation (also known as the ``virial 
method''; Kaspi et al. 2000; Greene \& Ho 2005b; see Peterson 2007, and 
references therein).  Based on such mass estimates, local AGNs do seem to obey 
roughly the same BH-host scaling relations as in inactive galaxies (Gebhardt 
et al. 2000b; Ferrarese et al. 2001; McLure \& Dunlop 2001; Nelson et al. 
2004; Onken et al. 2004; Greene \& Ho 2006b; Shen et al. 2008).  It is crucial 
to recognize, however, that at the moment this claim is based on {\it very}\ 
limited, and possibly biased, data.  The bright AGN core renders measurement 
of bulge luminosity and, in particular, central stellar velocity dispersion 
extremely challenging (see detailed discussion in Greene \& Ho 2006a).  
Notwithstanding these worries, many people now routinely assume that the 
BH-host scaling relations can be directly applied to AGNs, of arbitrarily high 
luminosity and redshift.

Here we propose a new angle to investigate the relationship between BH masses
and the host galaxies of AGNs.  In normal, inactive galaxies, it is well known
that the stellar velocity dispersion of the bulge tracks the maximum rotation
velocity of the disk, roughly as \vm\ = $(\sqrt{2}-\sqrt{3}) \times \sigma_*$ 
(Whitmore et al. 1979).  Although the theoretical 
underpinnings of this correlation are still murky, and the \vmsig\ relation 
is not as tight as has been claimed (Ferrarese 2002; Baes et al. 2003; 
Pizzella et al. 2005), nevertheless an empirical relation between \vm\ and 
\sig\ {\it does}\ exist (Courteau et al. 2007; Ho 2007a), which implies that 
\sig\ can be estimated from \vm.  Now, \vm\ can be measured straightforwardly 
through \hi\ observations for galaxies that are sufficiently gas-rich.  In the 
absence of a resolved rotation curve, \vm\ can be estimated from a single-dish 
measurement of the \hi\ profile, provided that we have some constraint on the 
inclination angle of the disk.  With the rotation velocity in hand, we can 
deduce immediately two physical quantities of interest: the total galaxy 
luminosity through the Tully-Fisher relation (Tully \& Fisher 1977), and, 
given an estimate of the size of the disk (e.g., from optical imaging), the 
dynamical mass of the system.  The advantages of this approach are clear.  
Since the \hi\ is distributed mostly at large radii (e.g., Broeils \& Rhee 
1997), it should be largely ``blind'' to the AGN core.  This allows us to 
circumvent the problems encountered in the optical, where attempts to 
disentangle the bulge from the active nucleus are maximally impacted.
Applying this principle, Ho (2007b) has recently used the integrated profile 
of the rotational CO line to infer certain properties of high-redshift quasar 
host galaxies.  
 
Apart from dynamical constraints, the \hi\ observations yield another piece
of information of significant interest---a first-order measurement of the cold
gas content.  As the raw material responsible for fueling not only the growth
of the BH but ultimately also the galaxy, one can hardly think of a more
fundamental quantity to ascertain.  How does the gas content of AGN host
galaxies compare with that in inactive galaxies?  Depending on one's view on
the evolutionary path of AGNs and the efficacy of gas removal by AGN feedback,
active galaxies may be more or less gas-rich than inactive galaxies.  Does
the gas content scale with the degree of nuclear activity?  Does it vary with
the nebular properties of the central source?  One of the major goals of this
paper is to try to address some of these basic questions, which to date
have been largely unanswered.

\section{The Database}

The analysis in this paper is based on the database presented in the companion
paper by Ho et al. (2008), to which the reader is referred for full details.
In brief, a comprehensive sample of 154 nearby type~1 (broad-lined) AGNs was 
assembled, consisting of new Arecibo\footnote{The Arecibo Observatory
is part of the National Astronomy and Ionosphere Center, which is operated by
Cornell University under a cooperative agreement with the National Science
Foundation.} observations of 101 sources with $z$ \lax\ 
0.11, mostly selected from the Fourth Data Release of the Sloan Digital Sky 
Survey (SDSS; Adelman-McCarthy et al. 2006), and a supplementary sample of 53 
other nearby sources collected from the literature.  In addition to basic \hi\ 
properties (line fluxes, line widths, and radial velocities), we also 
assembled optical data for the AGN (emission-line strengths, line widths, line 
centroid, nuclear luminosity) and the host galaxy (image, concentration index, 
size, axial ratio, total magnitude, central stellar velocity dispersion).  
From this material a number of important physical parameters were derived, 
including BH mass, Eddington ratio, morphological type, inclination angle, 
deprojected rotation velocity, \hi\ mass, dynamical mass, estimated host galaxy 
luminosity, and certain rudimentary properties of the narrow-line region.
Distance-dependent quantities were calculated assuming $H_0$ = 70 
\kms~Mpc$^{-1}$, $\Omega_{m} = 0.3$, and $\Omega_{\Lambda} = 0.7$. 

\section{Analysis}

\subsection{The \vmsig\ Relation in Active Galaxies}

To set the stage of using \vm\ as a surrogate dynamical variable to 
investigate BH-host scaling relations, we first examine the correlation 
between \vm\ and \sig\ for the active galaxies in our sample.  Including all 
objects that have measurements of both quantities, the scatter is 
discouragingly large (Fig.~1{\it a}).  Closer inspection, however, reveals 
that many of the extreme outliers correspond to objects that have potentially
untrustworthy deprojected rotation velocities because of uncertain inclination 
corrections, as well as a handful of sources whose central stellar velocity 
dispersions were estimated indirectly from the width of the \oii\ \lamb3727 
emission line (following Greene \& Ho 2005a).  Removing these objects improves 
the correlation.  As further discussed in \S 3.5, a sizable fraction of the 
objects in our sample contain non-classical \hi\ velocity profiles, which are 
either single-peaked, highly asymmetric, or both, indicative of strongly 
perturbed or dynamically unrelaxed \hi\ distributions.  If we further remove 
these cases---a cut that, unfortunately, drastically reduces the sample to 
only 40 objects---a much cleaner correlation between \vm\ and \sig\ emerges 
(Fig.~1{\it b}).  Because of the small number of objects and their limited 
dynamic range, the present sample is not well suited to define the \vm--\sig\ 
relation.  Nevertheless, the present sample still reveals a statistically 
significant correlation between \vm\ and \sig; the Kendall's $\tau$ 
correlation coefficient is $r = 0.53$, which rejects the null hypothesis 
of no correlation with a probability of $P$ = 98.2\%.  We overplot on the 
figure the \vm--\sig\ relation obtained from the 550 ``kinematically normal,'' 
inactive spiral galaxies from the study of Ho (2007a).  The ordinary 
least-square bisector fit for the inactive objects, 

\begin{equation}
\log \upsilon_m = (0.80\pm0.029) \log \sigma_* + (0.62\pm0.062), 
\end{equation}

\noindent
provides a reasonably good match to the AGN sample.  According to the 
Kolmogorov-Smirnov test, the probability of rejecting the null hypothesis that 
the AGN and non-AGN samples are drawn from the same parent population is $P$ = 
69.2\%.  We conclude that the two populations are not significantly different.
The scatter is still substantial, but recall that the zeropoint of the \vmsig\ 
relation depends on morphological type, such that early-type systems have a 
lower value of \vm/\sig\ than late-type spirals (\vm/\sig\ $\approx$ $1.2-1.4$ 
for E and S0, compared with \vm/\sig\ $\approx$ $1.6-1.8$ for spirals of type 
Sc and later; see Ho 2007a).  That the zeropoint for inactive spirals seems to 
roughly match the current sample of AGNs suggests that their host galaxies are 
mostly disk galaxies with modest bulge components, consistent with the actual 
estimated BH masses (median \mbh\ = $1.6\times10^7$ \solmass) and 
morphological types (median $T$ = 2.5, corresponding to Sab) of the objects.

Since local AGN host galaxies obey the \msigma\ relation, and we have just 
shown that they roughly follow the same 

\vskip 0.3cm
\figurenum{1}
\begin{figure*}[t]
\centerline{\psfig{file=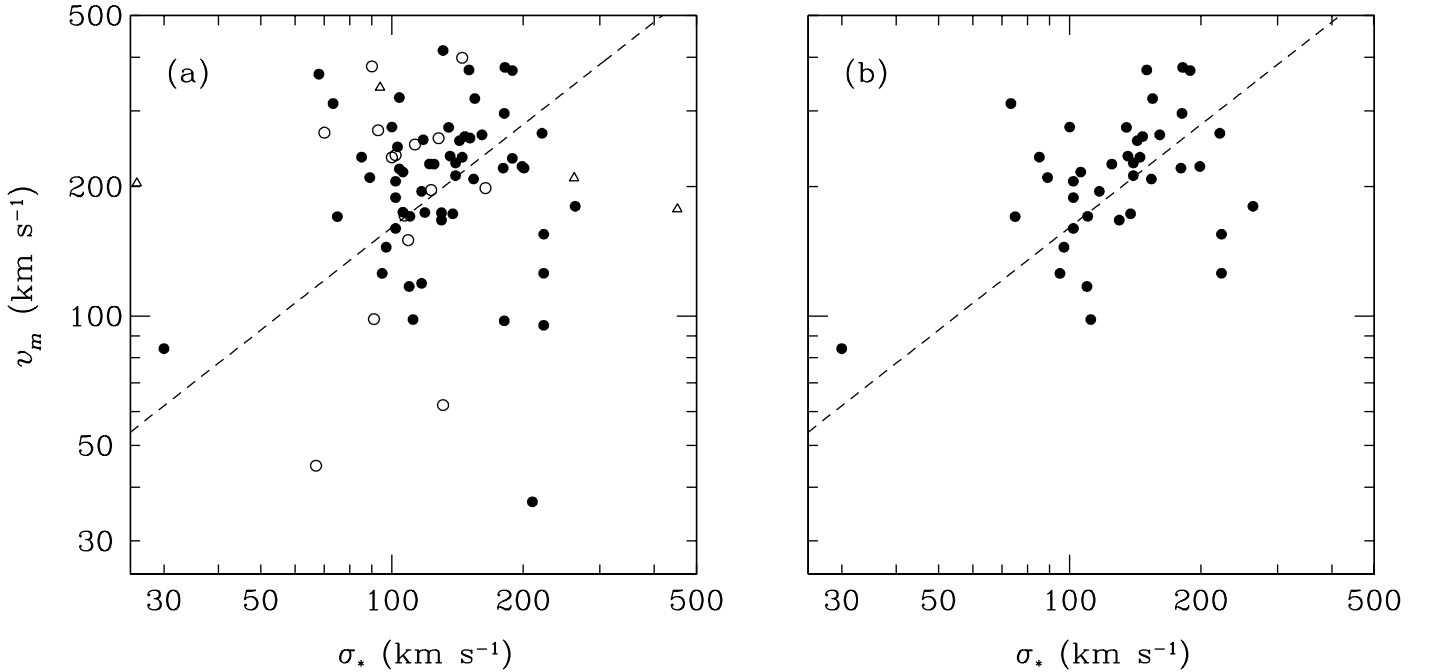,width=19.5cm,angle=-90}}
\figcaption[fig1.ps]{
Correlation between the central stellar velocity dispersion and the maximum
rotation velocity for the active galaxies in our sample.  Panel ({\it a})
plots all available data, including those that have uncertain inclination
corrections ({\it open circles}) and \oii-based estimates of the central
velocity dispersion ({\it open triangles}); objects with
both reliable rotation velocities and stellar velocity dispersions are plotted
as filled circles.  No selection has been done with regards to \hi\ profile
type.  Panel ({\it b}) excludes all objects with single-peaked and/or
asymmetric \hi\ profiles, as well as those with uncertain rotation velocities
and \oii-based estimates of the central velocity dispersion.  Note the
significant reduction in scatter.  The dashed line represents the fit to the
\vmsig\ relation for the 550 ``kinematically normal'' inactive spiral
galaxies from the sample of Ho (2007a); see text.
\label{fig1}}
\end{figure*}
\vskip 0.3cm

\noindent
asymmetric \hi\ profiles (panel {\it a})\footnote{For the rest of the 
$\upsilon_m-\sigma_*$ relation as in 
inactive galaxies, we expect \mbh\ to be correlated with \vm.  Not 
surprisingly, this is confirmed in Figure~2, which additionally reemphasizes 
that a tighter relation results after removing objects with single-peaked 
and/or paper, we have also excluded two objects from the literature sample that 
have suspiciously low values of \vm\ (Mrk 359, \vm\ = 38 \kms; Mrk 493, \vm\ = 
24.7 \kms).  Closer inspection of the original \hi\ data shows that Mrk 359 
has a highly unusual profile consisting of a narrow peak and a very broad base 
(Springob et al. 2005); moreover, the inclination angle given in Hyperleda, 
$i$ = 39.5\deg, seems inconsistent with the nearly face-on appearance of its 
Digital Sky Survey image (see Appendix in Ho 2007a for a discussion on the 
uncertainties associated with inclination angles listed in Hyperleda), 
suggesting that we have seriously underestimated the inclination correction.  
Mrk 493 may suffer from the same problem; the optical (SDSS) image of the 
object appears to be much more face-on than Hyperleda's value of $i$ = 
45\deg.}.  From the Kendall's $\tau$ test, $r = 0.58$ and $P$ = 99.9\%.
Overplotted on the figure is the relation expected from combining equation (1) 
with the \msigma\ relation for local AGNs obtained by Greene \& Ho (2006b),
$\log (M_{\rm BH}/M_\odot) = 4.02 \log (\sigma_*/200\,{\rm km~s^{-1}}) + 7.96$:

\begin{equation}
\log (M_{\rm BH}/M_\odot) = 5.1 \log \upsilon_m - 4.4.
\end{equation}

\noindent
As with Figure~1, we ran a Kolmogorov-Smirnov test to see whether the present 
sample of AGNs is drawn from the same population of objects used to 
define the predicted correlation between $M_{\rm BH}$ and \vm; the probability 
of rejecting the null hypothesis of no correlation is $P$ = 68.6\%.  Again, 
the two populations are not significantly different.

Nevertheless, the \mvm\ relation still 
contains significant scatter, even after omitting the kinematically peculiar 
objects.  What is responsible for this?  We believe that most of the scatter 
is intrinsic.  The virial BH mass estimates for nearby AGNs, based on broad 
H\al\ or H\bet, currently have an uncertainty of $\sim 0.3-0.5$ dex (Greene \& 
Ho 2006b; Peterson 2007), and errors associated with the inclination 
correction applied to the \hi\ line widths do not seem to be the dominant 
source of scatter (Fig.~2{\it b}).  Part of the scatter can be attributed 
to an effect related to Hubble type.  As mentioned above and 
discussed at length in Ho (2007a), \vm/\sig\ varies systematically with galaxy 
concentration or morphological type, and because our sample contains a wide 
range of morphological types (see Ho et al. 2008), we expect the zeropoint of 
the \mvm\ relation to be smeared by the actual mixture of galaxy types.  
This is illustrated in Figure~2{\it c}, wherein the sample is divided into 
four broad bins in Hubble type.  The small subsamples make it difficult to see 
the trend in detail, but comparison of the E+S0 galaxies with the later-type 
spirals clearly shows that the two subgroups are offset from each other, in 
the sense expected from the variation of \vm/\sig\ with morphological type.   

Taken at face value, another contribution to the scatter seems to be 
connected to variations in AGN properties, as shown in Figures~2{\it d} and 
2{\it e}, where we examine possible dependences on broad-line region line 
width and Eddington ratio.  Objects with broad H\al\ or H\bet\ full-width at 
half maximum (FWHM) $\leq 2000$ \kms, commonly designated ``narrow-line'' 
Seyfert 1 (NLS1) galaxies, have a slight tendency to be offset toward lower 
\mbh\ for a given \vm\ (Fig.~2{\it d}).  Since many NLS1 galaxies tend to have 
elevated accretion rates (e.g., Collin \& Kawaguchi 2004), the weak trend 
with FWHM may be more clearly discerned by dividing the sample according to 
Eddington ratio.  Indeed this expectation seems to hold, as shown in 
Figure~2{\it e}, where, within the limited statistics, there appears to be a 
monotonic decrease of \mbh/\vm\ with increasing \lbol/\ledd.  The trend with 
Eddington 

\vskip 0.3cm
\figurenum{2}
\begin{figure*}[t]
\centerline{\psfig{file=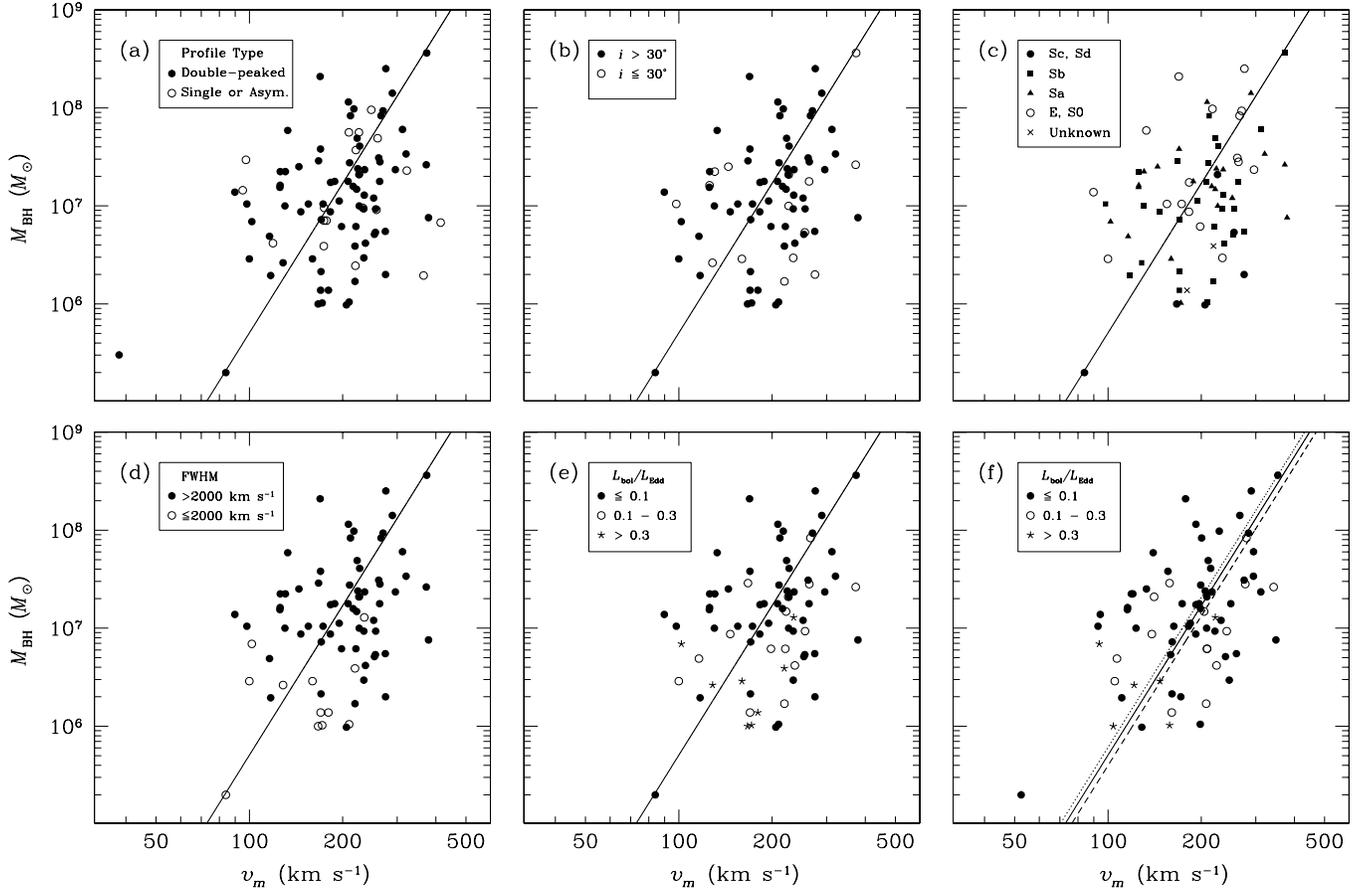,width=19.5cm,angle=-90}}
\figcaption[fig2.ps]{
Correlation between BH mass and the maximum rotation velocity deduced from
\hi\ observations.  The different panels test for possible dependences on
({\it a}) \hi\ profile type, ({\it b}) inclination angle correction, ({\it c})
Hubble type, ({\it d}) FWHM of broad H\al\ or H\bet, and ({\it e}) Eddington
ratio.  In panel ({\it a}), objects with uncertain inclination corrections
were removed; in panels ({\it b})--({\it e}), objects with single and/or
asymmetric \hi\ profiles were additionally removed.  Panel ({\it f}) is the
same as panel ({\it e}), but the different Hubble types have been adjusted to
a common zeropoint by shifting \vm.  The solid line represents the predicted
correlation for the whole sample, equation (2), given the \msigma\ relation of
Greene \& Ho (2006b) and the \vmsig\ relation (eq. 1); the dotted and
dashed lines show the zeropoint offset (in \mbh) for the objects with
$L_{\rm bol}/L_{\rm Edd} \leq 0.1$ and $L_{\rm bol}/L_{\rm Edd} > 0.1$,
respectively.
\label{fig2}}
\end{figure*}
\vskip 0.3cm

\noindent 
ratio is stronger than that with luminosity alone (not shown). Could 
the apparent variation of \mbh/\vm\ with FWHM and \lbol/\ledd\ be a secondary 
effect related to the dependence of \mbh/\vm\ on morphological type?  The most 
active AGNs in the nearby Universe live in moderate-mass, relatively late-type 
disk galaxies (e.g., Heckman et al. 2004; Greene \& Ho 2007b); these are 
precisely the hosts of high-\lbol/\ledd\ systems, such as NLS1s.   We attempt 
to remove the Hubble type dependence by shifting \vm\ for the different Hubble 
type subgroups to a common reference zeropoint (defined by the entire sample).
As expected, the scatter goes down, but, interestingly, the trend with 
\lbol/\ledd\ persists (Fig.~2{\it f}).

\subsection{Correlation between \mbh\ and Galaxy Dynamical Mass}

Although our \hi\ observations yield no information on the spatial distribution
of the gas, we can still obtain a rough estimate of the characteristic 
dynamical mass of the galaxy by combining the deprojected rotation velocity 
from the \hi\ line width with the optical diameter.  This approach is justified
for the following reason.  The size of the \hi\ disk in spiral galaxies over 
a wide range of Hubble types and luminosities tightly scales with the size 
of the optical disk; within 30\%--40\%, $D_{\rm H~{\tiny I}}/D_{\rm 25} 
\approx 1.7$ (Broeils \& Rhee 1997; Noordermeer et al. 2005), where 
$D_{\rm 25}$ is the optical isophotal diameter at a surface brightness level 
of $\mu_B = 25$ mag~arcsec$^{-2}$.  Thus, even if the absolute dynamical mass 
may be uncertain because of our ignorance of the size of the \hi-emitting 
disk, the {\it relative}\ masses should be reasonably accurate.  Following 
Casertano \& Shostak (1980), Ho et al. (2008) calculated dynamical masses for 
our AGN sample, which, interestingly, show a fairly strong correlation with 
BH mass (Fig.~3).  In particular, BH mass correlates more strongly with 
$M_{\rm dyn}$, which scales as $D_{\rm 25} \upsilon_m^2$, than with 
$\upsilon_m$ alone.  An ordinary least-squares bisector fit to the entire 
sample, after making the quality cut as above, gives

\begin{equation}
\log (M_{\rm BH}/M_\odot) = (1.29\pm0.11) \log (M_{\rm dyn}/M_\odot) - 
(6.96\pm1.17),
\end{equation}

\noindent
with an rms scatter of 0.61 dex.  The slope is formally steeper than unity, 
but this result should be regarded as provisional considering that most of the 
objects span a relatively narrow range of $M_{\rm dyn}$.  Closer inspection 
reveals that the zeropoint of the $M_{\rm BH}-M_{\rm dyn}$ correlation also 
depends on Hubble type (Fig.~3{\it a}).  At a given $M_{\rm dyn}$, early-type 
galaxies have a higher \mbh\ than late-type galaxies.  Relative to the 
best-fitting line for the whole sample, at a fixed log~\mbh\ the shift in 
$\log M_{\rm dyn}$ is $+0.19$, $-0.05$, $-0.10$, and $-0.59$ for galaxies of 
type E/S0, Sa, Sb, and Sc/Sd, 

\vskip 0.3cm
\figurenum{3}
\begin{figure*}[t]
\centerline{\psfig{file=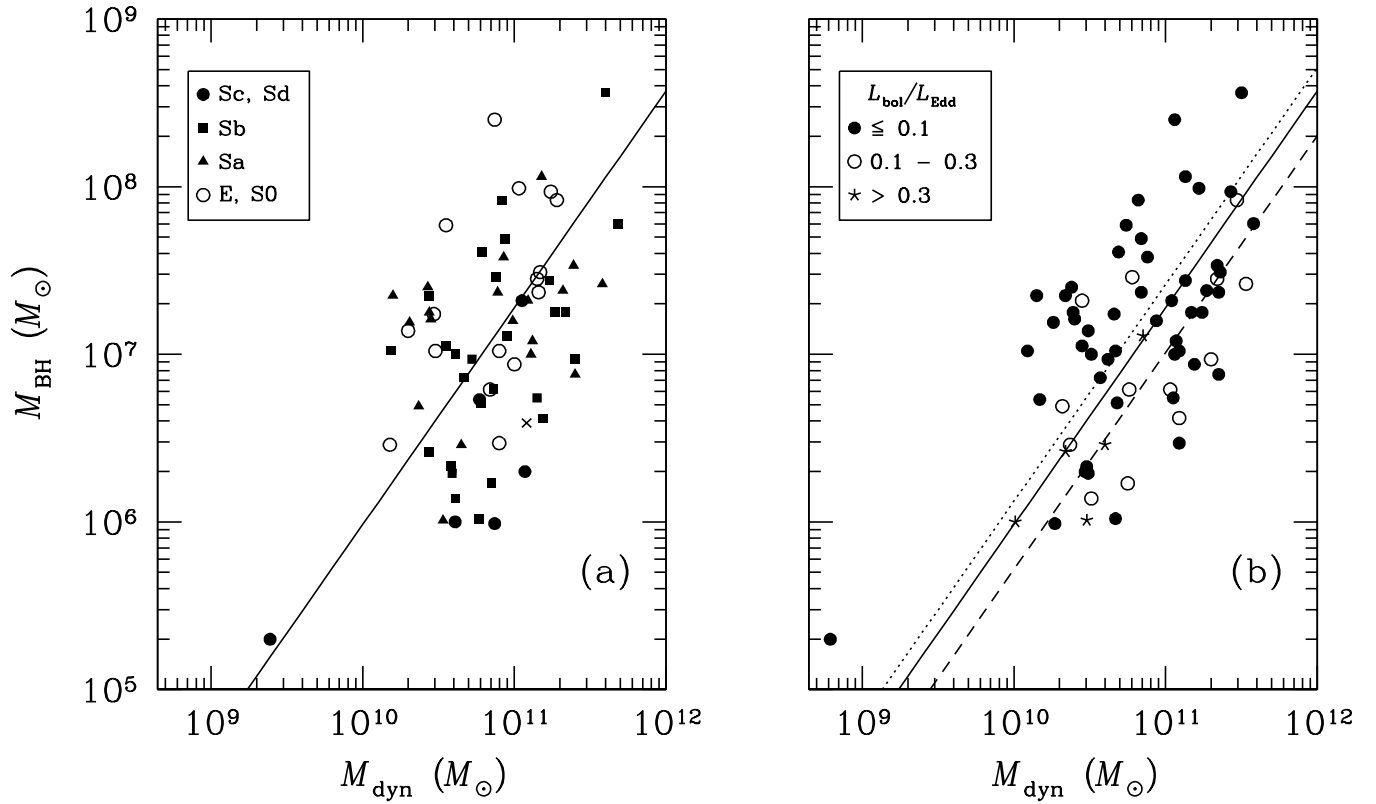,width=19.5cm,angle=-90}}
\figcaption[fig3.ps]{
Correlation between BH mass and dynamical mass of the host galaxy.  Objects
with uncertain inclination corrections and that have single and/or asymmetric
\hi\ profiles are excluded.  Panel ({\it a}) divides the sample into Hubble
types, and panel ({\it b}) bins the sample by Eddington ratio, after adjusting
the different Hubble types to a common zeropoint by shifting $M_{\rm dyn}$.
The solid line shows the ordinary least-squares bisector fit given in equation
(3) for the entire sample; the dotted and dashed lines show the zeropoint
offset (in \mbh) for the objects with $L_{\rm bol}/L_{\rm Edd} \leq 0.1$ and
$L_{\rm bol}/L_{\rm Edd} > 0.1$, respectively.
\label{fig3}}
\end{figure*}
\vskip 0.3cm

\noindent
respectively.  If we adjust the zeropoint of 
these groups by their corresponding offsets (in $M_{\rm dyn}$), the rms scatter 
of the resulting correlation decreases to 0.55 dex (Fig.~3{\it b}).  As in the 
case of the \mvm\ diagram, we find that at a given $M_{\rm dyn}$ objects with 
higher \lbol/\ledd\ tend to have smaller \mbh.

\subsection{The Tully-Fisher Relation for Active Galaxies}

The empirical correlation between \vm\ and total galaxy luminosity, first
introduced by Tully \& Fisher (1977), provides yet another avenue to assess
potential differences between active and inactive galaxies.  As described in 
Ho et al. (2008), approximate optical host galaxy luminosities can be obtained 
by comparing the integrated photometry with the nuclear spectroscopy.  Figure~4
plots the host galaxy absolute magnitude (after removing the AGN contribution) 
converted to the $B$ band, versus the maximum rotation velocity; overlaid for 
comparison is the corresponding Tully-Fisher relation for inactive (spiral) 
galaxies in the Ursa Major cluster (Verheijen 2001).  Considering first the 
entire sample (Fig.~4{\it a}), $M_{B,{\rm host}}$ loosely traces \vm, but 
the scatter is again considerable and the apparent correlation is formally
statistically insignificant ($r = -0.23$, $P$ = 93.3\%).  Pruning, as before, 
the objects with unreliable inclination corrections and kinematically peculiar 
profiles filters out most of the egregious outliers, such that the remaining 
sample falls mostly within the locus of inactive spirals (Figs.~4{\it b} and 
4{\it c}).  The formal correlation between $M_{B,{\rm host}}$ and \vm\ is 
still low ($r = -0.30$) and only marginally significant ($P$ = 95.0\%), 
probably because of the small sample size and limited dynamic range 
in \vm\ (90\% of the sample has \vm\ = 200$\pm$100 \kms).
Apart from the larger scatter exhibited by the active sample, an 
effect that can be attributed, at least partly, to its broad mixture of Hubble 
types, the approximate nature in which the host luminosities were estimated, 
and the sensitivity of the blue bandpass to extinction and stellar population 
variations (e.g., De~Rijcke et al. 2007), there are no other gross differences 
between the Tully-Fisher relation of active and inactive galaxies.  Different 
Hubble types define slightly offset, parallel sequences (Fig.~4{\it b}), an 
effect that has been well-documented in the $B$ band.  For a given $B$-band 
luminosity, an early-type spiral has a larger characteristic rotation 
amplitude than a late-type spiral (e.g., Rubin et al. 1985; De~Rijcke et al. 
2007).  To separate out this effect, we applied small corrections to the 
zeropoints (in host galaxy luminosity) of each Hubble type bin.  The adjusted 
distribution shows no obvious segregation by AGN properties, such as Eddington 
ratio (Fig.~4{\it c}).  

\subsection{\hi\ Content}

We next examine the \hi\ content of our sample, with special emphasis on 
whether AGN hosts differ in any noticeable way from inactive galaxies.  In 
absolute terms, our survey objects are quite gas-rich in neutral atomic 
hydrogen.  The \hi\ masses for the detected sources range from \mhi\
$\approx\,10^9$ to $4\times10^{10}$ \solmass\ (Ho et al. 2008).  Taking into 
account the upper limits for the nondetections, the Kaplan-Meier product-limit 
estimator (Feigelson 

\vskip 0.3cm
\figurenum{4}
\begin{figure*}[t]
\centerline{\psfig{file=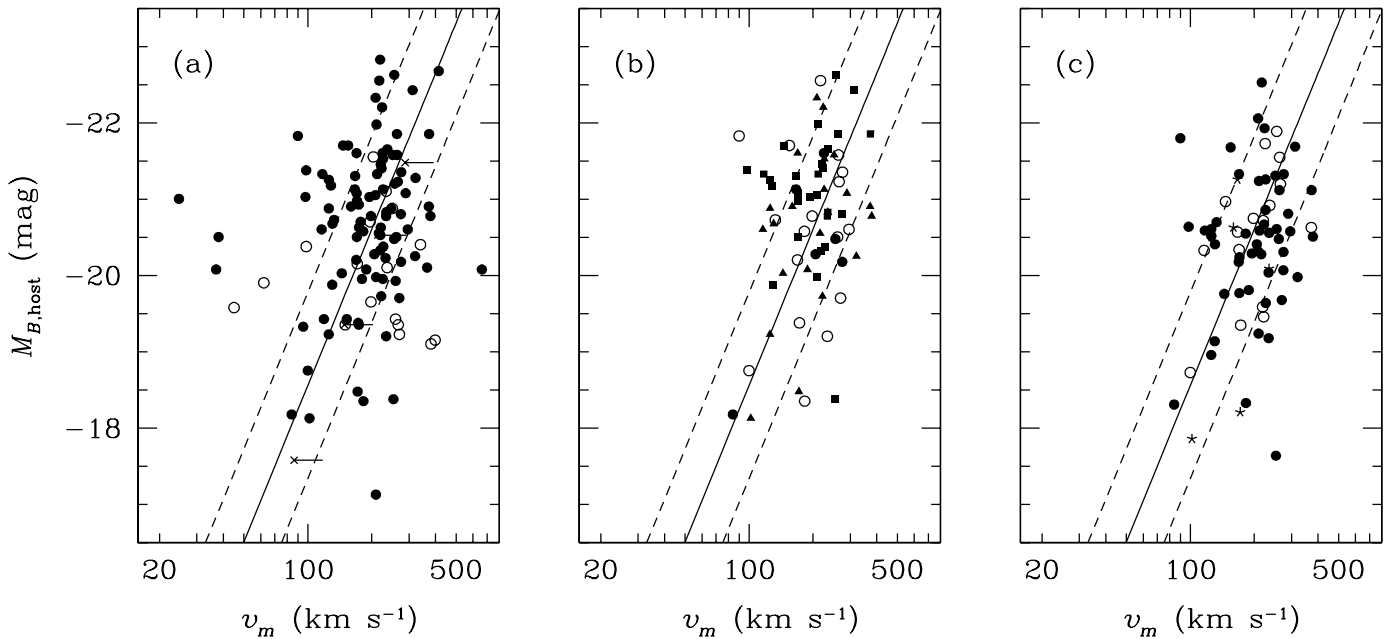,width=19.5cm,angle=-90}}
\figcaption[fig4.ps]{
The $B$-band Tully-Fisher relation for the active galaxies in our sample.
In panel ({\it a}), solid symbols denote reliable deprojected rotation
velocities, while less reliable values are marked with open symbols; crosses
indicate lower limits on \vm\ that lack estimates of the inclination angle.
In panel ({\it b}), the sample is divided according to Hubble type (open 
circles = E/S0; triangles = Sa; squares = Sb; filled circles = Sc/Sd), and
panel ({\it c}) bins the sample by Eddington ratio (solid, open, and star 
symbols denote objects with $L_{\rm bol}/L_{\rm Edd} \leq 0.1$, $0.1-0.3$, 
and $> 0.3$, respectively), after adjusting the different Hubble types to a 
common zeropoint by shifting $M_{B,{\rm host}}$.  Objects with uncertain 
deprojected rotation velocities and single-peaked and/or asymmetric \hi\ 
profiles were removed from panels ({\it b}) and ({\it c}).  The {\it solid}\ 
line represents the $B$-band Tully-Fisher relation for inactive spiral 
galaxies, as derived by Verheijen (2001) for galaxies in the Ursa Major 
cluster; the {\it dashed}\ lines mark the region that has twice the rms 
scatter of the Ursa Major sample.
\label{fig4}}
\end{figure*}
\vskip 0.3cm

\noindent
\& Nelson 1985) yields a mean of \mhi\ = 
$(7.0\pm0.66)\times10^{9}$ \solmass\ and a median of \mhi\ = 
$5.0\times10^{9}$ \solmass.   This is almost identical to the total 
\hi\ mass of the Milky Way ($5.5 \times 10^9$ \solmass; Hartmann \& Burton 
1997), and comparable to the average value for Sb spirals (Roberts \& Haynes 
1994).  The above statistics pertain just to the newly surveyed SDSS sources. 
The literature sample is both heterogeneous and possibly biased because the 
database from which the \hi\ data were compiled, 
Hyperleda\footnote{{\tt http://leda.univ-lyon1.fr/}}, does not report 
upper limits.  Nonetheless, we confirm that combining both samples does not
significantly alter these statistics.

The \hi\ content of galaxies varies greatly and systematically across the 
Hubble sequence (Haynes \& Giovanelli 1984; Roberts \& Haynes 1994).  Thus, 
in order to conduct a meaningful comparison of the \hi\ budget of active 
and inactive galaxies, we must know their morphological types.  Among the 
154 sources in our sample that have \hi\ detections or meaningful upper 
limits, 148 have estimates of both their morphological type and host 
galaxy optical luminosity.  Figure~5 shows the distribution of \hi\ masses
normalized to the $B$-band luminosity of the host galaxy, subdivided into six 
bins of Hubble types.  For reference, we computed the distribution of 
$M_{{\rm H~I}}/L_{B,{\rm host}}$ for 13,262 inactive galaxies culled from 
Hyperleda. These represent all galaxies in the database, reported to be 
current up to the end of 2003, that have reliable entries of \hi\ flux, 
morphological type, and $B$-band magnitude; known AGNs were excluded, as 
discussed in Ho et al. (2008; Appendix).  As previously mentioned, Hyperleda 
does not record upper limits for \hi\ nondetections, so these distributions 
should be viewed strictly as upper bounds.  Since the majority of the galaxies 
in Hyperleda are relatively bright and nearby, the detection rate of \hi\ 
among the spirals should be very high ($\sim$90\%; Haynes \& Giovanelli 
1984), such that the observed distribution can be regarded as being quite 
close to the true distribution.  The situation for early-type galaxies, 
however, is quite different, because the detection rate is only $\sim$15\% for 
ellipticals (Knapp et al. 1985) and $\sim$30\% for S0s (Wardle \& Knapp 1986), 
and so the distributions shown in the figure are highly biased.

Comparison of the active and inactive distributions reveals two interesting
points.  Among mid- to late-type spirals (Sb and later), the frequency 
distribution of $M_{{\rm H~I}}/L_{B,{\rm host}}$ is roughly the same for the 
two populations.  However, for the bulk of the sample, which comprise 
bulge-dominated Sa spirals and S0s, active galaxies appear to be {\it more}\ 
gas-rich than their inactive counterparts.  The most dramatic manifestation of 
this effect shows up among the ellipticals, although here we 
are handicapped somewhat by the small number of sources (9) in the active 
sample.  Note that among the E and S0 subgroups, the difference between 
the active and inactive samples is far greater than portrayed (although 
we have no rigorous statistical way to quantify this) because, as mentioned 
above, the inactive distribution is highly biased by the omission of upper 
limits.  (As noted, the literature data for the active sample are likely 
biased as well, but we verified that our conclusions are essentially 
unaffected if we exclude these data from the AGN sample.) The host galaxies of 
nearby type~1 AGNs across all Hubble types are at least as gas-rich as 
inactive galaxies, and among early-type systems their 

\vskip 0.3cm
\figurenum{5}
\psfig{file=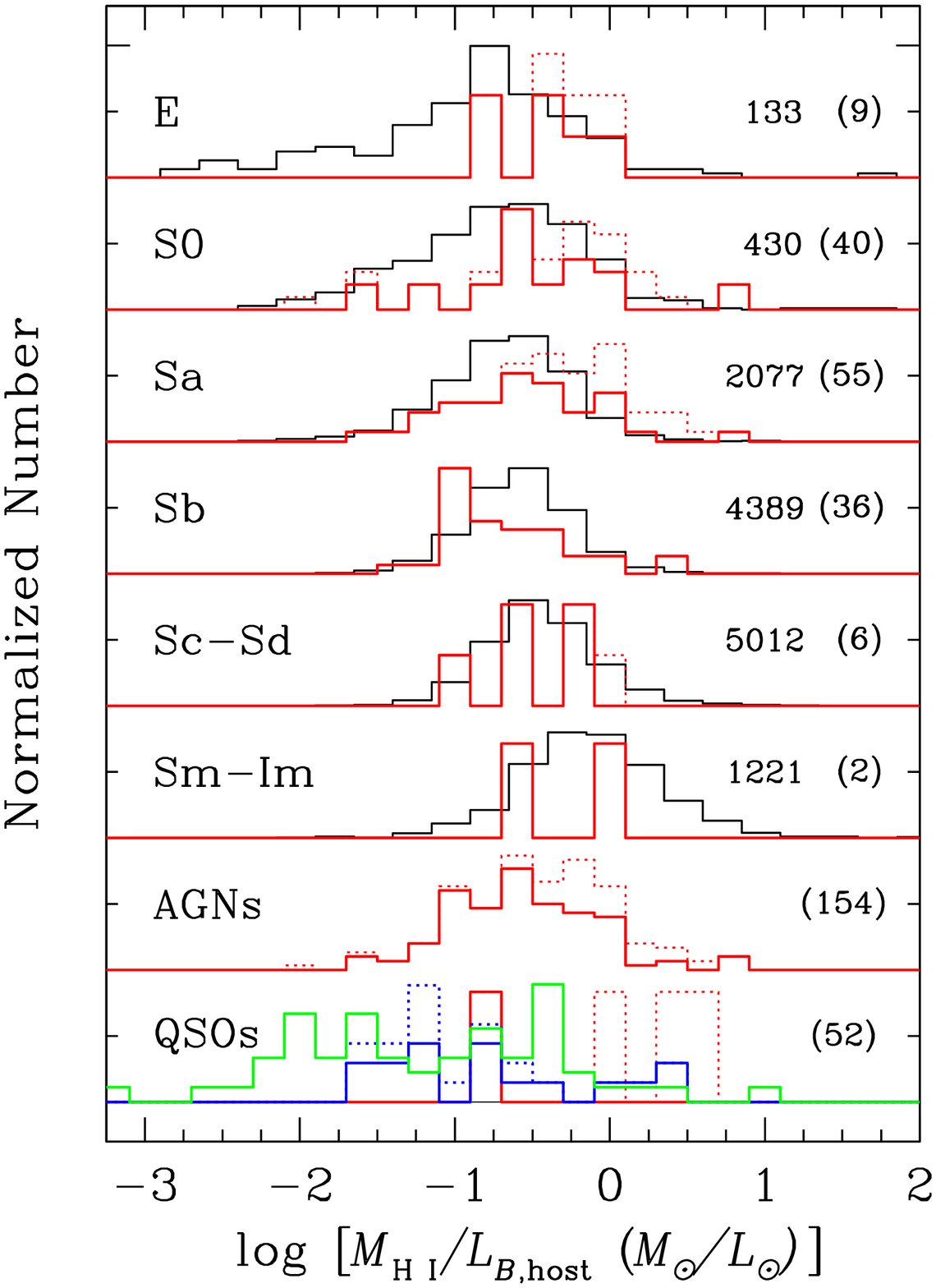,width=8.5cm,angle=0}
\figcaption[fig5.ps]{
The distribution of \hi\ masses normalized to the $B$-band luminosity of the
host galaxy, as a function of its Hubble type.  The black histograms show the
distributions for 13,262 inactive galaxies with \hi\ detections listed in
Hyperleda.  The AGN sample from our study is plotted in red histograms, with
upper limits indicated by dotted lines.  The entire sample of 154 objects is
plotted in the penultimate panel from the bottom, while the top panels show
the sample sorted by Hubble type.  The number of inactive galaxies in
each group is labeled, with the number of active objects shown in
parentheses.  The bottom-most panel isolates the quasars.  The red
histograms correspond to objects from the current \hi\ sample; the blue
histograms correspond to PG quasars observed in CO (Ho 2005a); the green
histograms correspond to $z < 0.5$ PG quasars with dust masses (Haas et al.
2003); see text for details.
\label{fig5}}
\vskip 0.3cm

\noindent
gas content appears to 
be markedly enhanced.

\subsection{Environmental Effects} 

Crude information on the spatial distribution or dynamical state of the gas 
can be ascertained from the degree to which the observed \hi\ line profiles 
deviate from the classical double-horned signature of an optically thin 
rotating disk.  Experience with nearby galaxies indicates that tidally 
disturbed systems often exhibit asymmetric, single-peaked, or otherwise highly 
irregular line profiles (e.g., Gallagher et al. 1981; Haynes et al. 1998).   
Given the modest signal-to-noise ratio of our data, Ho et al. (2008) decided 
not to implement a rigorous scheme to quantify the line morphology, but rather 
adopted a qualitative classification scheme in which obviously non-classical 
profiles were simply labeled as ``A'' (asymmetric), ``S'' single-peaked, or 
``AS'' (combination of both).  Among the 66 objects detected in our new survey, 
eight (12\%) are classified as ``A,'' five (8\%) are classified as ``S,'' 
and 16 (24\%) are classified as ``AS,'' for an overall frequency of 
non-classical profiles of 44\%.  Ho et al. (2008) performed a systematic 
census of physical neighbors within a search radius of 7\farcm5 around each 
object.  There seems to be no clear association between profile peculiarity 
and the presence of nearby neighbors.  While some objects with single-peaked 
and/or asymmetric profiles have plausible nearby companions, many do not;
at the same time, a number of objects with apparent companions exhibit 
seemingly regular line profiles.  The presence or absence of kinematic 
irregularity also appears to be uncorrelated with any of the global or AGN 
parameters that we have at our disposal.  We searched for possible differences 
in the following quantities, but found none that was statistically 
significant: morphology type, galaxy luminosity, total and relative \hi\ mass, 
AGN luminosity, broad-line region FWHM, BH mass, and Eddington ratio.

\subsection{Connections with AGN Properties}

The availability of optical data affords us an opportunity to explore possible
connections between the \hi\ and AGN properties of our sample.  Concentrating
on the SDSS objects, for which we have homogeneous optical spectroscopic 
measurements, we find that their narrow emission-line ratios place the 
majority of them in the territory of Seyfert nuclei (Fig.~6).  This is not 
surprising, since most of the SDSS-selected type~1 AGNs tend to have 
relatively high accretion rates (Greene \& Ho 2007b), which generally 
correspond to high-ionization sources (Ho 2004b).  The objects detected in 
\hi\ do not stand out in any noticeable way from the nondetections.  The 
same holds for the electron densities of the narrow-line region, as can be 
inferred from the line ratio \sii\ \lamb6716/\sii\ \lamb6731: the \hi\ 
detections are statistically indistinguishable from the \hi\ nondetections 
(Fig.~7).  Next, we searched for a possible dependence of AGN 
luminosity or Eddington ratio on \hi\ content, either in absolute (\mhi) or 
relative ($M_{{\rm H~I}}/L_{B,{\rm host}}$) terms, but again found 
none (Fig.~8).   There is, at best, a mild trend of increasing H\al\ luminosity 
with increasing \hi\ mass (Fig.~8{\it a}), but given the mutual dependence of 
the two quantities on distance, we regard this result as highly suspect.  

\section{Discussion}

\subsection{Black Hole-Host Galaxy Scaling Relations}

The relative ease with which integrated \hi\ emission can be detected in 
nearby AGNs opens up the possibility of using the global \hi\ line width as 
a new dynamical variable to investigate the scaling between BH mass and host 
galaxy potential.  A quantity that can be readily measured from single-dish 
spectra, the \hi\ line width has been widely used in a variety of 
extragalactic contexts as an effective shortcut to estimate \vm, the maximum 
rotation velocity of galaxy disks.  If \vm\ correlates with bulge velocity 
dispersion, \sig, as originally suggested by Whitmore et al. (1979), \vm\ can 
be used in place of \sig\ to define a \mvm\ relation to substitute for the more 
traditional \msigma\ relation. A number of authors have suggested that 
galaxies over a wide range in morphological types obey a tight correlation 
between \vm\ and \sig, to the point that \vm\ can actually replace, and 
perhaps should be regarded as more fundamental than, \sig\ (Ferrarese 2002; 
Baes et al. 2003; Pizzella et al.  2005).  This result has been challenged by 
Courteau et al. (2007) and Ho (2007a), who found, using much larger samples, 
that the ratio \vm/\sig\ shows significant intrinsic scatter and systematic 
variation across the Hubble sequence.  Nevertheless, a \vmsig\ relation 
{\it does}\ exist, at least statistically, and in circumstances when \sig\ is 
difficult or impossible to measure (e.g., in very bright or very distant 
AGNs), \vm\ may offer the best or possibly the {\it only}\ means of 
constraining the host galaxy.  Such was the motivation behind the recent study 
of Ho (2007b; see also 

\vskip 0.3cm
\figurenum{6}
\begin{figure*}[t]
\centerline{\psfig{file=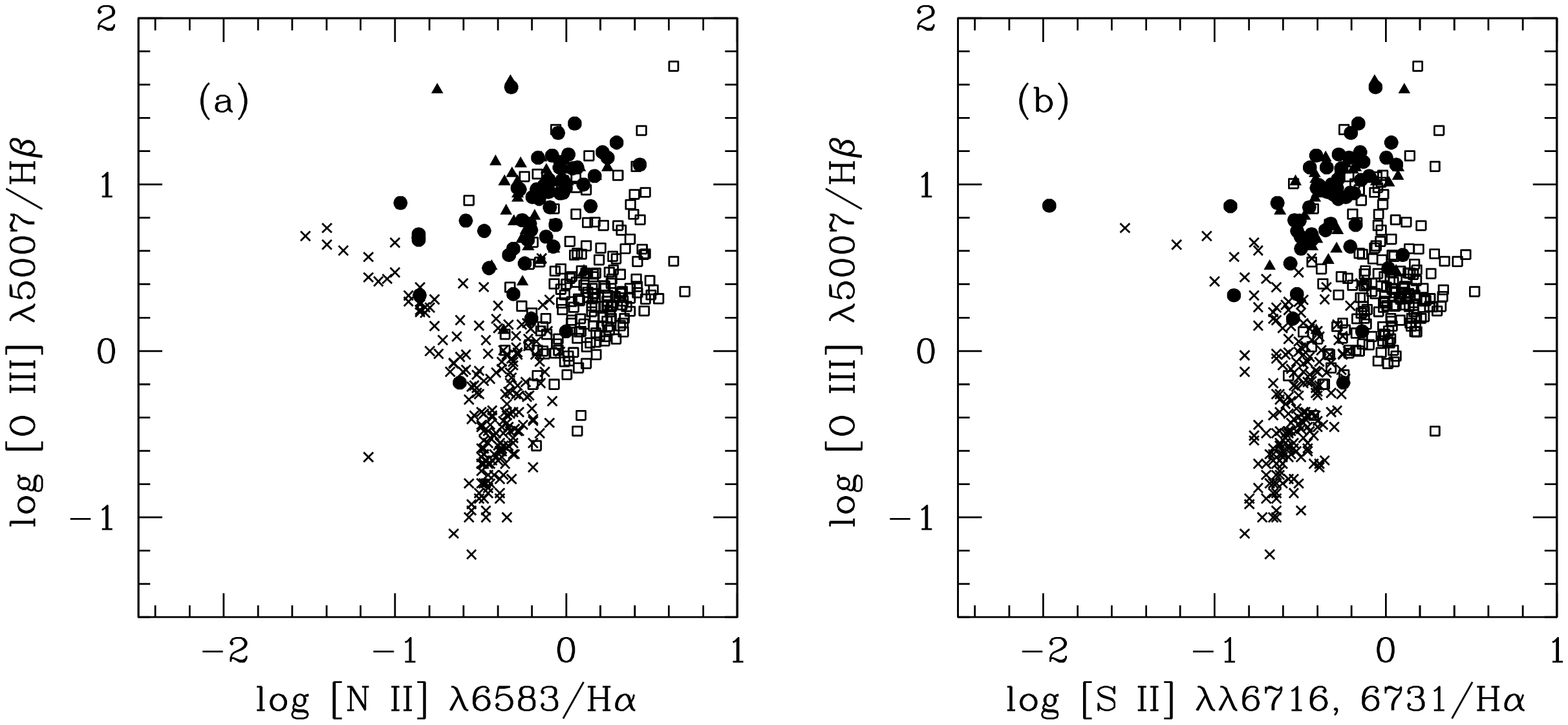,width=19.5cm,angle=-90}}
\figcaption[fig6.ps]{
The location of the \hi\ sample on the line intensity ratio diagrams
({\it a}) log \oiii\ \lamb5007/H\bet\ vs. log \nii\ \lamb6583/H\al\ and
({\it b}) log \oiii\ \lamb5007/H\bet\ vs. log \sii\ \lamb\lamb6716, 6731/H\al.
Objects detected and undetected in \hi\ are plotted as solid circles and
triangles, respectively.  The crosses represent \hii\ nuclei and the open
squares AGNs from the Palomar survey of nearby galaxies (Ho et al. 1997a).
The \hi\ sample mostly occupies the locus of Seyfert galaxies, with no 
apparent variation with \hi\ detection.
\label{fig6}}
\end{figure*}
\vskip 0.3cm

\noindent
Shields et al. 2006), who used the width of the 
rotational CO line, locally calibrated against \hi, to infer the masses of the 
host galaxies of high-redshift quasars.   The line width method, however, can 
prove to be useful even under less extreme conditions.  As discussed at length 
in Greene \& Ho (2006a), a variety of factors conspire to make measurement of 
\sig\ very challenging in galaxies containing bright AGNs, regardless of their 
redshift.  To bypass this difficulty, many studies resort to using gas velocity 
dispersions measured from narrow nebular lines, but this shortcut has its own 
set of complications (Greene \& Ho 2005a).   Others skip the kinematical route 
altogether and instead use the host galaxy (or bulge, if available) luminosity 
as the variable to relate to the BH mass.  However, cleanly 

\vskip 0.3cm
\figurenum{7}
\psfig{file=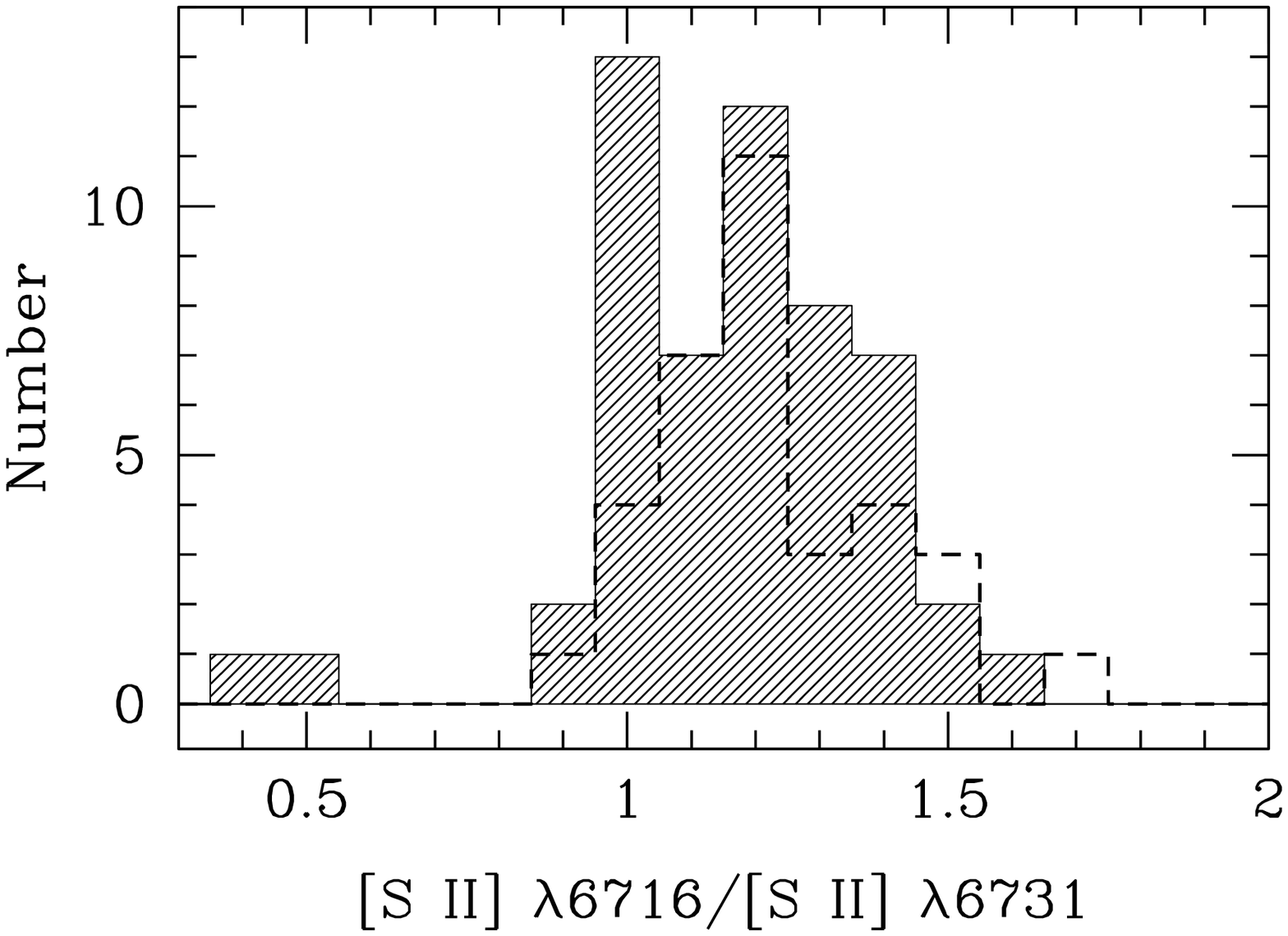,width=8.5cm,angle=0}
\figcaption[fig7.ps]{
Distribution of the line intensity ratio
\sii\ \lamb6716/\sii\ \lamb6731, which is inversely proportional to the
electron density, for objects detected (hashed histograms) and
undetected (open, dashed histograms) in \hi.
\label{fig7}}
\vskip 0.3cm

\noindent
decomposing the 
underlying galaxy, much less its bulge, in the presence of a bright AGN core 
is often a nontrivial task, even under the most ideal conditions (e.g., Kim 
et al. 2007).

Within this backdrop, we explored the correlation between BH mass and \vm.    
As expected, a loose correlation exists between these two quantities, but the 
scatter is enormous, $\sim 1$ dex (Fig.~2{\it a}).   Many of the extreme 
outliers correspond to objects with single-peaked and/or clearly asymmetric 
\hi\ profiles, while others have especially questionable inclination 
corrections; removing these produces a cleaner correlation.
Note that apart from being inefficient (roughly 40\% of the sample was 
rejected), this step imposes no serious complication, since classical 
double-horned line profiles are easy to recognize.  Nonetheless, even after 
this filtering, the scatter is still substantial (0.9 dex; Fig.~2{\it b}).  We 
identified a number of parameters associated with the AGN that seem to 
contribute to the scatter (FWHM of the broad emission lines, nuclear 
luminosity, and Eddington ratio), but an important contribution comes from
the morphological type variation within the sample (Fig.~2{\it c}).  As 
discussed in Courteau et al.  (2007) and Ho (2007a), the zeropoint of the 
\vmsig\ relation varies systematically with galaxy concentration (which is 
loosely related to galaxy morphology or bulge-to-disk ratio).  If we 
arbitrarily shift the zeropoints of the different Hubble type bins to the 
average relation for all Hubble types, the scatter decreases somewhat to 
$\sim 0.7$ dex (Fig.~2{\it f}).

More promising still seems to be the correlation between BH mass and 
galaxy dynamical mass, which requires knowledge of one additional parameter, 
namely the galaxy's optical diameter.  Retaining, as before, only the objects
with robust inclination corrections and kinematically normal \hi\ profiles, 
the $M_{\rm BH}-M_{\rm dyn}$ has an rms scatter of 0.61 dex, which further 
improves to 0.55 dex after shifting the different Hubble types to a common 
reference point (Fig.~3).  While this scatter is larger 

\vskip 0.3cm
\figurenum{8}
\begin{figure*}[t]
\centerline{\psfig{file=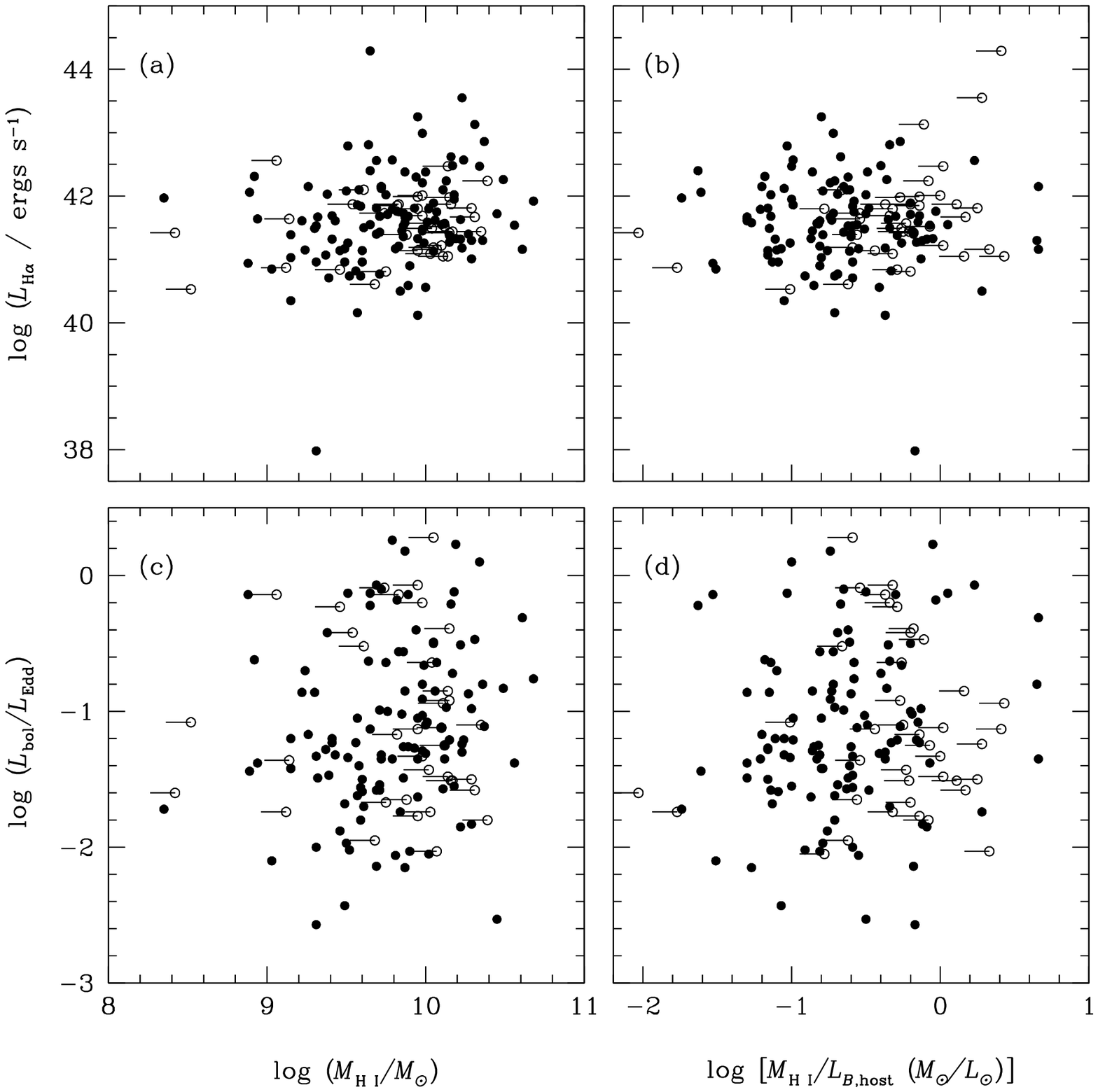,width=19.5cm,angle=0}}
\figcaption[fig8.ps]{
The variation of nuclear H\al\ luminosity (panels {\it a}\ and {\it b})
and Eddington ratio (panels {\it c}\ and {\it d}) with \hi\ content.  Objects
detected in \hi\ are plotted as filled symbols, while the nondetections are
marked as open symbols with arrows.
\label{fig8}}
\end{figure*}
\vskip 0.3cm

\noindent
than that of the 
local AGN \msigma\ relation (0.4 dex; Greene \& Ho 2006b), our estimates 
of the dynamical masses leave much room for improvement, both in terms of 
getting higher signal-to-noise ratio line profiles and deeper optical images, 
which will allow more accurate measurements of inclination angles and 
isophotal diameters and better estimates of morphological types.

\vskip 1.0cm
\subsection{The Chicken or the Egg}

One of the most intriguing results from our analysis is the tentative 
detection, both in the \mvm\ (Fig.~2{\it f}) and $M_{\rm BH}-M_{\rm dyn}$ 
(Fig.~3{\it b}) diagrams, of differential growth between the central BH and 
the host galaxy.  For a given host galaxy potential (\vm\ or $M_{\rm dyn}$), 
AGNs with higher accretion rates (Eddington ratios) have systematically 
{\it less}\ massive BHs, implying that in these systems the central BH, still 
vigorously accreting near its maximum rate, has yet to reach its final mass, 
which, from observations of inactive systems, we know to be a well-defined 
constant fraction of the total bulge luminosity or mass (Kormendy \& Richstone 
1995; Magorrian et al. 1998; H\"aring \& Rix 2004).  Many (but not all) of the 
objects with still-growing BHs turn out to have narrower broad Balmer lines 
(FWHM \lax\ 2000 \kms) because NLS1 galaxies tend to have higher accretion 
rates.  At a fixed \vm, the AGNs in our sample with \lbol/\ledd\ \gax\ 0.1 on 
average have BHs 0.19 dex (factor of 1.5) less massive than those with 
\lbol/\ledd\ \lax\ 0.1; in terms of fixed $M_{\rm dyn}$, the difference in BH 
masses between the low- and high-accretion rate subgroups is $\sim$0.41 
dex (factor of 2.6).  In their study of type 1 AGNs with stellar velocity 
dispersion measurements, Greene \& Ho (2006a) also noticed that the 
best-fitting \msigma\ relation for AGNs has a small zeropoint offset of 
$-$0.17 dex relative to Tremaine et al.'s (2002) fit of the \msigma\ relation 
for inactive galaxies.  A result similar to that of Greene \& Ho (2006a) has 
been reported by Shen et al.  (2008), who additionally note that the amplitude 
of the \msigma\ relation for AGNs depends on Eddington ratio, in the same 
sense that we find in our study.  The overall qualitative agreement among the 
three independent studies lends credence to the idea that the most highly 
accreting BHs are still actively growing.

Mathur et al. (2001; see also Grupe \& Mathur 2004; Mathur \& Grupe 2005), 
using the width of the \oiii\ \lamb 5007 line as a substitute for \sig, a 
shortcut that allows large numbers of AGNs to be placed on the \msigma\ 
relation, proposed that NLS1s contain BHs that are undermassive with respect 
to their bulges, by as much as an order of magnitude.  Employing the same 
technique, Bian \& Zhao (2004) arrived at a similar result, as did Wandel 
(2002), who used bulge luminosities as a reference.  The claim that NLS1s 
are ``young'' AGNs, however, has not been universally embraced, in part 
because different authors, using the same methods, have arrived at conflicting
conclusions (e.g., Wang \& Lu 2001), because of worries concerning the 
reliability of the \oiii-based velocity dispersions (Botte et al. 2004; Greene 
\& Ho 2005a), and, most seriously, because direct measurements of stellar 
\sig\ in NLS1s find no compelling evidence that they deviate from the \msigma\ 
relation (Barth et al. 2005; Botte et al. 2005; Greene \& Ho 2006b).  Part of 
the confusion stems from the very definition of the NLS1 class.  Strictly 
speaking, NLS1s are classified on the basis of optical spectroscopic criteria 
(e.g., Pogge 2000), which, apart from the (fairly arbitrary) line width limit 
of FWHM $\leq$ 2000 \kms, are neither rigorously defined nor universally 
followed.  That many cataloged and well-studied NLS1s have high accretion 
rates (e.g., Pounds et al. 1995; Collin \& Kawaguchi 2004) is a statistical 
reflection of the fact that bright AGNs in the local Universe tend to have 
moderately low-mass BHs (Heckman et al. 2004; Greene \& Ho 2007b), and that 
soft X-ray selection preferentially favors systems in their high-state (Greene 
\& Ho 2007a).  There is no requirement that NLS1s must have high accretion 
rates.  The present sample, in fact, provides two useful examples.  By their 
line widths alone, both NGC 4051 and NGC 4395 qualify as NLS1s, and yet the 
former has \lbol/\ledd\ = $6.3\times10^{-2}$ while the latter, with 
\lbol/\ledd\ = $2.7\times10^{-3}$, formally has the lowest Eddington ratio in 
the sample.

That \mbh/\vm\ and \mbh/$M_{\rm dyn}$ decrease {\it systematically}\ with
increasing \lbol/\ledd\ suggests not only that the growth of the BH and the
growth of the galaxy are not finely synchronized, but also that there is a
preferential sequence to their coevolution: during any particular common
growth event, star formation {\it precedes}\ BH growth.  Thus, at least for 
the local sample under consideration, we have a tentative solution to the 
proverbial ``the chicken or the egg problem.''  Phenomenologically, starburst 
activity comes first, followed by AGN activity---not the other way around, and 
probably not concurrently.  This resembles the scenario envisioned by Ho 
(2005a, 2005b; see also Kim et al. 2006) in his analysis of the star formation 
rate in nearby quasars.  Interestingly, the sequence of events seems to be 
reversed at high redshifts, where the best evidence to date suggests that 
galaxy growth actually {\it lags}\ BH growth (Peng et al. 2006a, 2006b; Shields
et al. 2006; Ho 2007b).  One might speculate that the pathway for BH growth 
and its interdependence on the host galaxy could be radically different for 
luminous quasars in massive, early-type galaxies, compared to the much more 
modest AGNs under consideration here, which reside mostly in later type, disk 
systems (e.g., Hopkins \& Hernquist 2006).  On the other hand, we note that 
moderate-redshift ($z=0.36$) AGNs, mostly Seyfert galaxies not too dissimilar 
from those in our sample, also follow the trend of high-redshift quasars in 
showing a higher ratio of BH mass to host galaxy (bulge) mass (Woo et al. 
2006; Treu et al. 2007).  We cannot, at the moment, reconcile these 
conflicting results, which urgently need to be verified.

We end the discussion on BH growth with some cautionary remarks.  Throughout 
this study we take at face value that the virial BH masses contain no major 
sources of systematic error, other than the $\sim$0.5 dex uncertainty in the 
zeropoint (Greene \& Ho 2005b; Peterson 2007).  As discussed in Greene \& Ho 
(2007c; Appendix), however, our assumption that the virial mass formalism is 
invariant with respect to AGN properties is, at the moment, more an article of 
faith than an indisputable fact.  The size-luminosity relation on which the 
virial formalism depends continues to undergo revision and is especially poorly 
constrained at the low-luminosity end (Bentz et al. 2006).  The very structure
of the broad-line region, and hence the geometric (``$f$'') factor of the 
virial relation, may depend on accretion rate or Eddington ratio, but 
presently we have no reliable means to quantify this effect (Greene \& Ho 
2007c).  Strictly speaking, therefore, the trend of decreasing \mbh/\vm\ and 
\mbh/$M_{\rm dyn}$ with increasing \lbol/\ledd\ can be interpreted as evidence 
that \mbh\ is systematically underestimated in high-\lbol/\ledd\ AGNs.  
Resolving this degeneracy lies beyond the scope of this work.  At the 
moment we also cannot prove that the identified trends are not the result of 
subtle mass-dependent selection effects.  In their investigation of the local 
BH mass function of nearby AGNs, Greene \& Ho (2007b) note that low-mass BHs, 
which even at their Eddington limit are still relatively faint AGNs, 
preferentially make it into a magnitude-limited survey such as SDSS if they 
reside in more luminous (more massive) galaxies.  

\subsection{Host Galaxies and Environment}

Even though most sizable nearby galaxies harbor central BHs, they exhibit 
very feeble traces of nuclear activity (Ho et al. 1997b).  Only a tiny 
fraction of the local galaxy population accrete at rates high enough to 
qualify them as respectable AGNs (Ho 2004b; Greene \& Ho 2007b).  Why is this?
The low space density of luminous AGNs in the local Universe surely reflects 
some combination of the overall reduced gas supply in present-day galaxies 
as well as their more placid dynamical environments, but precisely which
physical parameters---internal or external to the system---actual trigger the 
onset of nuclear activity in any given galaxy remains largely an unsolved 
problem (Ho et al. 1997c, 2003; Ho 2004b).  In a similar vein, and especially
given the recent discussion of the purported link between BH and galaxy 
growth, it is of interest to know what impact AGN activity actually has on the
properties of the host galaxy.  

Our new \hi\ survey offers an opportunity to examine these issues from some 
fresh angles.  Despite the limited quality of the data, our visual examination 
the optical images do not give the impression that the galaxies or their 
immediate environments are particularly unusual.  Some clear cases of 
interacting or binary systems exist within a separation of 50 kpc (Ho et al. 
2008; see Fig.~3{\it g}, SDSS~J165601.61+211241.2; Fig.~4{\it b}, 
SDSS~J130241.53+040738.6), but overall the optical morphologies of the hosts
appear undisturbed and large (i.e. nearly equal-mass) physical companions are 
rare.  There are also no discernible differences between the objects 
detected and undetected in \hi\ (cf. Fig.~3 vs. Fig.~4 in Ho et al. 2008).

Excluding the objects with anomalous \hi\ profiles, the rest of the sample is 
not just morphologically normal but also dynamically normal.  We draw this 
inference from our analysis of the Tully-Fisher relation 
for the active galaxies, which, besides the anticipated larger scatter due 
to various measurement uncertainties, reveals no striking differences compared 
to regular disk galaxies (Fig.~4).  As an aside, it is worth making an obvious 
point: the fact that our objects obey the Tully-Fisher relation implies that 
the \hi\ must be roughly regularly distributed, at least regular enough so
that its integrated line width traces the flat part of the galaxy rotation 
curve.  Hutchings et al. (1987) and Hutchings (1989) previously failed to 
see a clear Tully-Fisher relation for their sample of AGNs because of the 
high frequency of asymmetric \hi\ profiles.  By contrast, the more extensive 
study by Whittle (1992) found that Seyfert galaxies do define a Tully-Fisher 
relation, but one that is offset toward lower velocities compared to normal 
spiral galaxies.  He interpreted this as an indication that Seyfert galaxies 
have lower mass-to-light ratios (by a factor of $1.5-2$), possibly as a 
result of enhanced star formation.  Our results do not agree with Whittle's.  
We have verified that the discrepancy between the two studies does not stem 
from the different fiducial relations chosen for inactive spirals.  While 
Whittle's reference Tully-Fisher relation has a much steeper slope than the 
one we adopted (from Verheijen 2001), the two zeropoints are very similar.  We 
speculate, but cannot prove, that the low-velocity offset seen by Whittle may 
be caused by contamination from objects with kinematically peculiar, 
preferentially narrow \hi\ profiles.  Inspection of our data (Fig.~4) reveals 
that prior to excluding the objects with asymmetric or single-peaked profiles 
our sample also has a mild excess of low-velocity points.  Indeed, a 
nonnegligible fraction of all galaxies, irrespective of their level of nuclear 
activity, show this behavior, which Ho (2007a) attributes to dynamically 
unrelaxed gas acquired either through minor mergers or primordial accretion.  

Enhanced star formation frequently seems to precede nuclear activity (Kauffmann 
et al. 2003; see discussion in Ho 2005b), but at least for the AGN luminosities
probed in this study, the extent of the young stellar population has left no 
visible imprint on the mass-to-light ratio of host galaxies, insofar as 
we can gauge from the Tully-Fisher relation.

Do AGN host galaxies possess a higher fraction of kinematically anomalous 
\hi\ profiles than inactive galaxies?  Our study, as in previous ones 
(Mirabel \& Wilson 1984; Hutchings et al. 1987), tentatively suggests that the
answer is yes, but a quantitative comparison is difficult.  Single-dish 
surveys of isolated, (mostly) inactive disk galaxies also report frequent 
detections of asymmetric \hi\ line profiles (e.g., Baldwin et al. 1980; Lewis 
et al.  1985; Matthews et al. 1998), which are attributed to noncircular 
motions, unresolved companions, or genuine perturbations in the spatial 
distribution of the gas. Richter \& Sancisi (1994) and Haynes et al. (1998), 
for example, find asymmetric profiles in roughly 50\% of the spiral galaxies 
they surveyed.  Similarly, Lewis (1987) studied a large sample of face-on 
spiral galaxies and found narrow \hi\ profiles to be rare.  He interprets this 
result to imply that a substantial fraction of the galaxies must have 
distorted \hi\ distributions, whose large-scale motion is misaligned with 
respect to the plane of the stellar disk.  Thus, in absolute terms the 
frequency of non-classical \hi\ profiles in our sample (44\%; \S3.5) is very 
similar to the frequency reported for inactive galaxies, but we are wary to
draw any firm conclusions from this because the signal-to-noise ratio of our
spectra is generally much lower than those of the control samples, and 
because of the qualitative nature of our profile classification.  

The kinematically anomalous objects account for 17\% of the 792 galaxies studied 
by Ho (2007a), but this fraction depends on Hubble type, being more common 
among early-type systems.  Among S0 galaxies, the fraction reaches 40\% (Ho 
2007a), nearly identical to the frequency found among the active systems 
studied here, which, although strictly not all S0 galaxies, nonetheless tend 
to be bulge-dominated disk galaxies.  As in the case of inactive galaxies 
(Ho 2007a), the kinematic peculiarity for the AGN sample cannot be attributed 
to tidal interactions or comparable-sized nearby neighbors. 

\vspace{0.3cm}
\subsection{Gas Content and Implications for AGN Feedback Models}

Insofar as their \hi\ content is concerned, the host galaxies of the AGNs in 
our sample are endowed with plenty of gas.  The most meaningful metric is the 
specific rather than the absolute gas mass, and because this quantity spans
a wide range across the Hubble sequence (Roberts \& Haynes 1994), we can 
sharpen the comparison even further by specifying the morphological type of
the galaxy.  This exercise (Fig.~5) convinces us that type~1 AGNs are {\it at 
least}\ as gas-rich as inactive galaxies, and among early-type hosts, 
beginning with Sa spirals and certainly by the time we reach S0s and Es, AGN 
hosts appear to be even {\it more}\ gas-rich than their inactive counterparts.
For example, the sensitive survey of 12 inactive E and S0 galaxies by Morganti 
et al. (2006) detected \hi\ emission in nine objects, of which only one has 
$M_{{\rm H~I}}/L_{B,{\rm host}} > 0.1$.  By contrast, among the 30 active E 
and S0 galaxies in our survey that were detected in \hi, 26 (87\%) have 
$M_{{\rm H~I}}/L_{B,{\rm host}} > 0.1$.  Our results are qualitatively 
consistent with those of Bieging \& Biermann (1983) and Mirabel \& Wilson 
(1984).  Because of the nature of the control sample of inactive objects (only 
detections are available), we hesitate to be more specific, but E and S0 
galaxies hosting AGNs may be overabundant in \hi\ by as much as an order of 
magnitude.   Previous studies of nearby elliptical and S0 galaxies have hinted 
of a possible statistical connection between \hi-richness and radio activity 
(Dressel et al. 1982; Jenkins 1983; Morganti et al. 2006), two well-known 
examples being the low-power radio galaxies NGC~1052 (Knapp et al. 1978) and 
NGC~4278 (Raimond et al. 1981).  The validity of our conclusion, of course, 
critically depends on the accuracy of the morphological types.  However, to 
bring the average $M_{{\rm H~I}}/L_{B,{\rm host}}$ ratios of the E and S0 
subgroups to conform to the average values of inactive galaxies, we would have 
to shift the morphological types to as late as Sb or even Sc spirals.  It 
seems unlikely that the morphologies could be so blatantly wrong.  
Alternatively, perhaps we have grossly miscalculated the host galaxy optical 
luminosities.  Without improved optical imaging (\S5), however, these 
alternative explanations remain purely hypothetical.  

The global gas content bears no relationship to the level of AGN activity, 
either in terms of absolute luminosity or Eddington ratio (Fig.~8).  Ho et al. 
(2003) had come to the same conclusion for nearby LINERs and Seyferts, but now 
the same holds for AGNs on average 2--3 orders of magnitude more luminous.  
Whether \hi\ is detected or not also makes no impact whatsoever on the optical 
spectrum (Figs.~6 and 7).  Perhaps none of this should come as a surprise. 
The \hi\ data, after all, probe spatial scales vastly disproportionate to those 
relevant for the AGN central engine.  Moreover, the accretion rates required
to power the observed activity are quite modest.  The H\al\ luminosities of 
our sample range from $\sim 10^{40}$ to $10^{44}$ \lum, with a median value of 
$4\times 10^{41}$ \lum.  Adopting the H\al\ bolometric correction given in 
Greene \& Ho (2007b) and a radiative efficiency of 0.1, this corresponds 
to a mass accretion rate of merely $\sim 2\times 10^{-2}$ \solmass\ \peryr.  

Many current models of galaxy formation envision AGN feedback to play a 
pivotal role in controlling the joint evolution of central BHs and their host 
galaxies. During the major merger of two gas-rich galaxies, each initially 
seeded with its own BH, gravitational torques drive a large fraction of the 
cold gas toward the central region of the resulting merger remnant.  Most of 
the gas forms stars with high efficiency in a nuclear starburst, at the 
same time feeding the central BH at an Eddington-limited rate.  This process 
consumes a large amount of the original cold gas reservoir.  The combined 
energy generated from supernova explosions and the central engine wrecks havoc 
on the rest of the interstellar medium in the galaxy, shocking it to high 
temperatures and redistributing it to large scales, thereby shutting off 
further star formation and accretion.  While the specific formulation of the 
problem and the methodology for solving it may vary from study to study (e.g., 
Granato et al. 2004; Springel et al. 2005; Croton et al. 2006; Hopkins et al. 
2006; Hopkins \& Hernquist 2006; Sijacki et al. 2007; Di~Matteo et al. 2008), 
one generic prediction runs constant: the onset of nuclear activity, 
especially during the peak phase of accretion when the central object unveils 
itself as an optically visible AGN, in concert with supernova feedback from 
the accompanying starburst, liberates so much energy on such a short timescale 
that the bulk of the cold gas gets expelled from the galaxy.  None of the 
existing models makes very precise predictions about the cold gas content 
during the evolution of the system, but it seems clear, regardless of the 
details, that during the optically visible (unobscured) phase of the AGN the 
host galaxy should be {\it deficient}\ in cold gas.  This expectation is 
inescapable if AGN feedback is to have as dramatic an effect on the evolution 
of the host galaxy as has been suggested.

Our observations present a challenge to the framework of AGN feedback just 
outlined.  Far from being gas-deficient, the host galaxies of optically 
selected type~1 AGNs are, if anything, unusually gas-rich, and at the very 
least we can state with confidence that their gas content is normal.  To be 
fair, the vast majority of our sample consists of Seyfert galaxies.  They are 
hosted in disk (S0 and spiral) galaxies, which probably never experienced many 
major mergers, and their nuclei fall far short of luminosity threshold of 
quasars, which, although rare, dominate the BH mass density in the Universe 
(e.g., Yu \& Tremaine 2002).  These objects probably fall outside of the 
purview of major merger-driven models (Hopkins \& Hernquist 2006).  Our 
sample, on the other hand, {\it does}\ contain seven objects that formally 
satisfy the luminosity criterion of quasars, of which four (PG~0844+349, 
PG~1426+015, PG~2130+099, and RX~J0608.0+3058) are detected in \hi.   All four 
are very gas-rich, with \hi\ masses ranging from \mhi\ = $5 \times10^9$ to 
$2\times10^{10}$ \solmass\ and $M_{{\rm H~I}}/L_{B,{\rm host}}$ values that 
are not obviously low.  

Beyond the present set of observations, two other lines of evidence contradict 
the notion that optically selected quasars are gas-deficient.  Scoville et al. 
(2003) performed an unbiased search for CO emission from all $z < 0.1$ quasars 
from the ultraviolet-selected sample of Palomar-Green (PG) sources (Schmidt \& 
Green 1983), finding that the majority of them ($\sim$75\%) contain abundant 
molecular gas, ranging from $M_{{\rm H_2}} \approx 10^{9}$ to $10^{10}$ 
\solmass; those that were undetected have upper limits consistent with the 
detections.  Bertram et al. (2007) report very similar statistics from their 
recent CO study of low-luminosity quasars selected from the Hamburg/ESO 
survey.  These molecular gas masses are comparable to, if not greater than, 
those typically found in mid- to late-type spirals.  Ho's (2005a) compilation 
of CO observations of other PG quasars further reinforces this point, and even 
more extreme molecular gas masses ($M_{{\rm H_2}} \approx 10^{10}-10^{11}$ 
\solmass) have been detected in high-redshift quasars (Solomon \& Vanden~Bout 
2005, and references therein).  The Milky Way, for reference, has 
$M_{{\rm H_2}} = 2\times10^9$ \solmass\ (Scoville \& Good 1989).  Of the three 
PG quasars that overlap with our \hi\ sample, for instance, PG~1426+015 has 
$M_{{\rm H_2}} = 6.5\times10^9$ \solmass, PG~2130+099 has $M_{{\rm H_2}} = 
3\times10^9$ \solmass, and PG~0844+349 has an upper limit of $M_{{\rm H_2}} = 
1\times10^9$ \solmass.  Combining this with the \hi\ masses, all three objects 
have a total (\hi\ $+\,{\rm H}_2$) neutral gas mass in excess of $10^{10}$ 
\solmass.  Yet another way to recognize that PG quasars have plenty of cold 
gas comes from noting that substantial amounts of dust have been detected in 
them.  Haas et al. (2003) appraised the dust content in a sample of 51 PG 
quasars by combining far-infrared continuum observations along with millimeter 
and submillimeter data.  Their derived dust masses range from $M_{{\rm dust}} 
=7\times10^5$ to $3\times10^8$ \solmass, with a median value of $8.8\times10^6$ 
\solmass.  For a Galactic molecular gas-to-dust ratio of 150, this corresponds 
to $M_{{\rm H_2}} = 1.3\times10^9$ \solmass; for a higher gas-to-dust ratio of 
600 (Young \& Scoville 1991), $M_{{\rm H_2}} = 5.3\times10^9$ \solmass.  These
indirect estimates of molecular gas mass are consistent with those derived 
from CO observations.

To graphically compare the gas masses of PG quasars side-by-side with those 
of the lower luminosity AGNs in our sample, we converted the molecular and 
dust masses into approximate \hi\ masses.  We adopt, for the sake of 
illustration, $M_{{\rm H_2}}$/\mhi\ = 3, a value characteristic of early-type 
spirals (Young \& Scoville 1991), and $M_{{\rm H_2}}/M_{{\rm dust}} = 600$.
We calculated BH masses for the sources using the H\bet\ line widths from 
Bororon \& Green (1992) and optical continuum luminosities from the 
spectrophotometry of Neugebauer et al. (1987), and then inferred $B$-band host 
galaxy luminosities using the correlation between BH mass and bulge luminosity 
given in Kormendy \& Gebhardt (2001), for simplicity assigning all the 
luminosity to the bulge.  As summarized in the bottom panel of Figure~5, PG 
quasars span almost 4 orders of magnitude in $M_{{\rm H~I}}/L_{B,{\rm host}}$, 
from objects as gas-poor as the most extreme ellipticals to as gas-rich as the 
most gas-dominated dwarf irregulars.  Although this exercise becomes 
progressively more uncertain for high-redshift objects, we can attempt a 
similar order-of-magnitude calculation for the nine high-redshift quasars 
listed in Ho (2007b), using BH masses listed therein and the molecular gas 
masses given in Solomon \& Vanden~Bout (2005).  Under the same assumptions 
adopted for the PG quasars, we find an average value of 
$\log \langle M_{{\rm H~I}}/L_{B,{\rm host}} \rangle \approx -0.9$, in the 
middle of the range observed for PG quasars (Fig.~5).  In fact, this estimate, 
which assumes a local value for the ratio of BH mass to bulge luminosity, is 
most likely a lower limit because high-redshift quasars 
in general (Peng et al. 2006a, 2006b) and this subset of objects in particular 
(Shields et al. 2006; Ho 2007b) seem to exhibit a higher ratio (by a factor of 
a few) of BH mass to host galaxy mass.  

Given the lack of specific theoretical predictions, we cannot attempt a 
quantitative comparison of our results with AGN feedback models.  Nonetheless, 
we wish to stress that the typical unobscured quasar possesses a gas reservoir 
normal for its stellar content and does {\it not}\ appear to be depleted in 
cold gas.  Whether this poses a difficulty or not for AGN feedback models 
deserves further consideration.

\section{Future Directions}

The pilot study presented here can be improved and extended in several 
ways.  A number of the \hi\ observations would benefit from longer 
integrations.  Roughly $1/3$ of our sample was undetected in the current 
experiment, and the upper limits on their \hi\ masses remain relatively high;
nearly half of the nondetections have limits above \mhi\ $=\,10^{10}$ 
\solmass. It would be useful to improve these limits by a factor of a few, 
down to \mhi\ $\approx 5\times 10^{9}$ \solmass, the scale of the Milky Way.
Higher signal-to-noise ratio spectra of the detected sources would allow us 
to better study the \hi\ line profiles, especially to quantify the apparent 
excess in line asymmetry and its implications for the dynamical state of 
the gas and the circumgalactic environment of the host.  In terms of new 
observations, it would be particularly worthwhile to observe a larger sample 
of high-Eddington ratio systems in order to verify the trends with Eddington 
ratio identified in this study.  This can be easily accomplished using the 
current SDSS database (Greene \& Ho 2007b, 2007c).  To put more stringent 
constraints on AGN feedback scenarios, it would be highly desirable to 
increase the number of high-luminosity AGNs (quasars) in the survey.  Given 
the rarity of nearby quasars and the redshift limit of Arecibo, this will be a 
difficult task, but some progress can still be made.   

The current project represents a major initiative in characterizing the cold 
gas content in type~1 AGNs.  We have demonstrated that \hi\ observations can 
yield many useful insights into the nature of AGNs, BHs, and their host 
galaxies.  This tool should be applied to other classes of AGNs.  A useful 
extension of this program would be to observe a well-matched sample of type~2 
(narrow-line) AGNs in order to test their relationship to type~1 objects.  Kim 
et al. (2006; see also Ho 2005b and Lal \& Ho 2007), for example, have argued 
that type~2 quasars are the evolutionary precursors, as opposed to simply 
misoriented counterparts, of type~1 quasars.  If true, type~2 sources are 
likely to be more gas-rich than their type~1 counterparts, or perhaps their 
interstellar medium might be less dynamically relaxed, differences that can 
potentially be discerned through \hi\ observations.   

The analysis presented here critically depends on a number of quantities 
derived from optical images of the host galaxy, namely its inclination angle, 
isophotal diameter, luminosity, morphological type, and, to a lesser extent, 
local environment.  All of these measurements can be greatly improved 
with the help of better optical imaging.  In many instances, the SDSS images 
(Ho et al. 2008) simply lack the necessary depth or resolution to yield 
clear-cut measurements of these parameters.  High-resolution (e.g., 
{\it Hubble Space Telescope}) images would be particularly valuable, as they 
would simultaneously yield a reliable decomposition of the AGN core from the 
host galaxy, as well as a quantitative measurement of the bulge component. 
All of the correlations presented in this study should be reexamined once 
such data become available, to see if their scatter can be reduced.

We have stressed the utility of \hi\ observations, but, of course, a full
inventory of the cold interstellar medium should include the molecular 
component as well, which, depending on the galaxy type, can dominate over the 
atomic component.  A systematic CO survey of the objects in our \hi\ program 
would be highly desirable, not only to complete the gas inventory but also to 
bootstrap the kinematical analysis to the CO line width.  Unlike the current 
limitations of \hi\ observations, the rotational transitions of CO can be 
observed to almost arbitrarily high redshifts (Solomon \& Vanden~Bout 2005), 
a necessary ingredient to access AGN host galaxies at early epochs.  Ho 
(2007b) discusses the use of the CO Tully-Fisher relation as a powerful 
tool to investigate the host galaxies of high-redshift quasars.

\acknowledgements
The work of L.~C.~H. was supported by the Carnegie Institution of Washington
and by NASA grants HST-GO-10149.02 and HST-AR-10969 from the Space Telescope
Science Institute, which is operated by the Association of Universities for
Research in Astronomy, Inc., for NASA, under contract NAS5-26555.  Support for
J.~D. and J.~E.~G. was provided by NASA through Hubble Fellowship grants
HF-01183.01-A and HF-01196, respectively, awarded by the Space Telescope
Science Institute.  We thank Paul Martini for useful correspondence and 
the anonymous referee for helpful suggestions.

\end{document}